# Investigation of cluster structure of light nuclei H, He and Li isotopes, in three-body final states photodisintegration processes in the energy range 50-250 MeV using photon beams of Yerevan Synchrotron

N. Demekhina, H. Hakobyan, A. Sirunyan and H. Vartapetian

Alikhanyan Yerevan Physics Institute, Yerevan, Armenia

**Abstract**

In this work we propose to investigate the cluster structure of the excited states of nuclei produced in the photodisintegration reactions on $^6$Li and $^7$Li targets, with three bodies (clusters and nucleons) in final state using bremsstrahlung and linear polarized beams. The analysis of the experimental data with the use of Dalitz diagrams allows to investigate 16 cluster structures of 9 isotopes of H, He and Li nuclei in four reactions of photodisintegration with the above-mentioned targets.

**Introduction**

Many experimental and theoretical works have been performed to study the light nuclei structures. The structure of the excited states of these light nuclei is a subject of increasing interest and is widely discussed in the modern theoretical analyses[1], that corresponds to a possible existence and manifestation of cluster structure inside of these nuclei.

The experimental studies of the excited states of light nuclei are performed by different methods within possibilities of existing experimental facilities. In the majority of experiments on investigation of the light nuclei structure, ion (π-meson) and photon (electron) beams are used. Some available experimental information published after 1995 and 2000 is briefly presented below.

**1. Ion (π-meson) beams**

The experimental results concern the excited states of the isotopes $^6$He, $^6$Be, $^6$Li, $^7$Li and $^{11}$B including their structure.

**1.1 Ion beams**

In reaction $^6$Li ($^7$Li, $^7$Be) at energy 93 MeV two broad levels at energies $E_x$ =14MeV and 25 MeV were revealed [2]. However, the $^6$He excited states decay modes were not determined in this experiment. Reaction $^6$Li ($^7$Li,$^7$Be)t at energy 65 MeV was investigated [3]. The breakup of the new resonance state $^6$He with energy $E_x$= 18.0 ±0.5 MeV and width Γ=7.7±1.1 MeV into two tritons was observed. The contribution of this breakup channel consists of (90±10)%.

Reaction $^6$Li ($^7$Li,$^7$Be,t)x at energy 455 MeV also via breakup $^6$He into two tritons also at $E_x$=18.0±1 MeV with Γ=9.5 ±1 MeV was investigated [4]. In the case of nucleus $^6$Be in reaction $^6$Li ($^3$He,$^3$He,t)$^3$He at energy 450 MeV the breakup into two $^3$He at level energy $E_x$=18.0 ±1.2 MeV and

$\Gamma$=9.2 ±1 MeV was registered [4]. The two breakup channel contribution (for $^6$He and $^6$Be) formed (70±10)%.

Reaction $^7$Li ($^3$He,α) at energy 450 MeV was investigated [5]. The two body breakup process of $^6$Li→t+$^3$He via a broad resonance state at $E_x$=21MeV and $\Gamma$=21 ±2 MeV was observed. This resonance state in $^6$Li was divided off line into two excited cluster states (t+$^3$He) at energies $E_x$=18.0±0.5 MeV with $\Gamma$=5.0 ±0.5 MeV and $E_x$=22.0±1 MeV with $\Gamma$=8 ±1 MeV .

Reaction $^{11}$B (d, d') at energy 200 MeV was investigated [6]. The results of measuring the differential cross sections of this reaction are in good agreement with the results of some model analysis, taking into account the (2α+t) cluster structure for the excited state of $^{11}$B at $E_x$=8.56 MeV.

### 1.2 π-meson beams

Reaction $^9$Be ($\pi^-$, 2t)x with $\pi^-$ meson beam of Meson Factory in Los Alamos (USA) by means of two tritons coincidence system was measured [7]. The extraction of the three tritons breakup channel in the investigated reaction using Dalitz diagram was made. The obtained results are pointing to the three excited states in $^6$He, decaying via two tritons at $E_x$=15.8 ±0.6 MeV , $\Gamma$=1.1 ±0.6 MeV , $E_x$=20.9 ±0.3 MeV , $\Gamma$=3.2 ±0.5 MeV and $E_x$=31.1 ±1 MeV , $\Gamma$=6.9 ±2.3 MeV.

Concerning the results of the above-mentioned investigations: we can state that in the investigations performed after 2000 the obtained results differed even in case of more studied isotope $^6$He concerning the quantity of excited states, their energy and width.

### 2. Photon beams (bremsstrahlung, coherent polarized photon- CB, tagged photons)

### 2.1 Linear –polarized beams (CB)

In photoproduction reactions on isotopes $^6$Li and $^7$Li in energy up to 100 MeV the results concerning the cluster structure $^7$Li (α+t) and $^6$Li [($^3$He +t) ,( d+α)] were obtained. Experimental data including differential cross section (dσ/dΩ) and cross section asymmetry (Σ) with bremsstrahlung and polarized photon beams are measured.

Reaction γ+$^6$Li→$^3$He+t with CB beams at energies 25-70 MeV was investigated [8]. The experimental results on Σ value are in good agreement with the model predictions, where the ground state in $^6$Li has cluster structure α+d.

It should be mentioned also that photodisintegration experiments γ+d →n+p and γ+$^6$Li→p+n+X, were performed on Yerevan Synchrotron with CB beam in the energy range 300-900 MeV[9]. The experimental results show that Σ-values in both reactions, measured at θ* =90°, coincide in the limits of the statistical errors. These results do not contradict the cluster structure in $^6$Li ground state (α+d).

In reactions γ+$^6$Li→$^3$He +t and γ+$^7$Li→$^4$He +t with real and polarized photons [10] the measured Σ values are in good agreement with theoretical analysis results including cluster structures $^3$He +t and $^4$He +t. At the same time dσ/dΩ differed from model predictions.

**2.2 Two-body reaction investigation with tagged photons** (MAX-Lab Lund University, Sweden)

Photodisintegration reaction $^6$Li(γ, cc') with tagged photons (intensity γ ≈ 6x10$^5$ γ/s.MeV) was investigated at electron energy equal to 100 MeV by using tagged photons method coincidence (cc', charged particles) [11]. The yields of the coincidence measurements for pairs: pt, pd, dt, d$^3$He and t$^3$He are obtained .

Reaction $^6$Li (γ,p) was measured at average tagged photon energies of 59 and 75 MeV [12]. The results of the analysis have shown that most of the observed strength is apparently due to the three-body breakup channels.

In two-body reaction γ+$^6$Li→$^3$He +t, measured with tagged photons in the energy range 50-80 MeV[13], the dσ/dΩ values are in contradiction with model prediction for the cluster structure ($^3$He +t) of $^6$Li ground state [10].

In 2006 an experiment [14] was proposed on Lund electron accelerator in MAX-Lab. The reactions γ+$^6$Li→$^3$He +t and γ+$^7$Li→α +t had to be investigated in the energy range 40-100 MeV with polarized and bremsstrahlung photons. The aim of this experiment was to study the cluster structure of $^6$Li and $^7$Li nuclei. Unfortunately, the information about the above-mentioned experiment has not been available.

**3. The method of the cluster structure investigation**

This method is based on the study of the reaction of photodisintegration light nuclei A= $^6$Li and $^7$Li in three selected particles in the final state (γ+A→1+2+3). These three particles are mainly two clusters and one nucleon. Taking into account the studied problem a coincidence is realized between two particles (from three), mainly two clusters, among the following possible four cluster-particles: H(p),$^2$H(d), $^3$H(t), $^3$ He.

The advantage of the above-mentioned method compared to the method of reaction with two particles in the final state γ+A → 1+2 is the possibility to investigate not only the features of the excited state of the stable Li targets on two clusters (or nucleon +cluster), but to investigate the excited states of the unstable isotopes breaking up into two known particles (nucleon +cluster or two clusters) in accordance with the scheme γ + A → (1,2)* + 3 with the decay of the excited isotopes (1,2)* → 1+2 and others. In our case, where A= $^6$Li, $^7$Li, it is possible to investigate not only the excited states of target-isotopes $^6$Li, $^7$Li, but also the excited states of the following isotopes $^3$H, $^4$H, $^5$He, $^6$He and $^5$Li (see below).

**3.1 The realization of the method**

For the fulfilment of these investigations, except the knowledge of the incident photon energy $E_\gamma$ (see below), two conditions for photon beams are necessary:

- the beam should consist of bremsstrahlung (not tagged) photons (with $E_{\gamma Max}$ up to π-meson production threshold). This will enable to use high intensity beams (≥$10^{10}$γ/s) in the experiment, as well as to get polarized photon beams CB in crystals up to energy $E_\gamma$ ≈100 MeV (polarization $P_\gamma$≥50%).
- the extracted photon beam should have a good time stretching (>2-3 ms). Under high intensity this will enable to diminish the number of accidental coincidences.

From the above-mentioned it follows that it is necessary to restore the energy of the incident photon beam ($E_\gamma$) for each event of three-body photodisintegration reaction (A) - γ+A → 1+2+3. For this purpose:

1) one of the three particles which is not registered (particle 3) should be a nucleon (p, n);

2) the other two particles-clusters, 1 and 2 should be stable at production, detection and identified, all their three kinematics parameters (p(T), θ, φ) should be measured in coincidence.

Within the fulfillment of these two conditions, for the two considered targets

A ($^6$Li, $^7$Li) and four registered particle-clusters (particles 1,2: $^1$H(p), $^2$H(d), $^3$H(t),$^3$He), taking into account the two well-known conservation laws (of charge and baryon number) in reactions, the following four possible three-body reactions γ + A → 1+2+3 are obtained (particle 3 is not registered):

$$\gamma + {}^6Li \to t + d + p$$
$$\gamma + {}^6Li \to {}^3He + d + n$$

$$\gamma + {}^7Li \to t + t + p$$
$$\gamma + {}^7Li \to {}^3He + t + n$$

**3.2 Cluster structures of the excited isotopes states. Dalitz diagrams.**

Experimental investigation of the above-mentioned four reactions is carried out by means of the registration of two particles (1) and (2) in each of the studied reactions using two telescopes in coincidence (or ≈ 4π detector). Each telescope has a known threshold for the particle registration for their kinetic energy and allows to determine the mass-charge of particles (the kind of particle (1), (2), its kinetic energy ($T_i$) and the emitting direction (2 angles $\theta_i$, $\varphi_i$), where i=1-4(p,d,t, $^3$He).

Knowledge of these parameters for each of the two particles in four reactions will enable to restore the cluster structure of the excited isotopes participating in the reactions (i.e. two target-isotopes and produced isotopes in four reactions). It is possible to establish the number and type (structure) of the produced isotopes in these reactions.

After the interaction of photons with the known target (A) and the registration of the six parameters of the two particles (1) and (2) in the given reaction, $E_\gamma$ incident photon energy and the effective mass of the excited state are established, i.e. the energy of the excited target-isotope disintegration in the system of three particle-clusters (1, 2, 3).

There are two channels for disintegration of the excited states of isotope-targets:
- a statistical channel (γA) → 1+2+3, which represents a physical background channel of the reaction (γ + A) and which can be calculated as three-body phase space.
- a channel of production and decay of the excited isotope states in the given reaction γ + A → 1+2+3 in the form of three two-body processes γ +A → (12)+3; (13)+2; (23)+1 with three resonance-molecules (12)*,(13)*,(23)*. They are decaying in the final three-particle state.

Dalitz diagrams

These diagrams have enabled to show "resonances" existence (excited hadron states: nucleon, meson) in three-body reactions with elementary particles. These diagrams can also be used to investigate the photodisintegration reactions in three particle-clusters and reveal the production of clusters (molecular resonance) in the reactions γ+A→1+2+3.

The calculations in the frames of Dalitz diagrams for every case are made in the common center mass system (CM) (γA)=(123). For this procedure it is necessary to know kinetic energy ($T_i$) and angle ($\theta_i$) values for these three particles in CM system for γA reaction. These circumstances enable to calculate three diagrams in two coordinate system $T_1 =f(T_2)$ , $T_2 =g(T_3)$ and $T_3 =h(T_1)$ for every reaction. The contours of these diagrams and effective masses of three possible resonances ($M_{12}$, $M_{13}$, $M_{23}$) created in reactions γA→ $M_{12}$+3; $M_{13}$+2; $M_{23}$+1 are also calculated. It may also be shown that the distribution of the ($T_i$,$T_j$) points on the diagrams surface is uniform for phase space of the three-body reaction. For this kind of reactions the spectra of the effective masses for two particles from (1,2,3) may be calculated.

### 3.3 The results of Dalitz diagrams analysis

From the four reactions with $^6$Li,$^7$Li targets (see 3.1) cluster structures for separated isotopes may be obtained:

For $^6$Li target there are the following 1,2,3 stable particles in two reactions γ + $^6$Li:

[t,d,p], [$^3$He,d,n].

Cluster structures can be of two types: "cluster+particle" and "cluster+cluster".

For excited $^6$Li target can be observed :

cluster+particle: ($^5$He+p), ($^5$Li+n)

two clusters ($^3$He+t), ( $^4$He + d)

For excited isotopes products in the reaction γ+ $^6$Li:

cluster+particle: $^3$H(d+n), $^3$He (d+p), $^4$He[(t+p),($^3$He+n)]

two clusters $^5$He (t+d), $^5$Li ($^3$He+d)

The number of cases: "cluster + particle" is equal to six and for two clusters it is equal to four, the total sum - ten.

For $^7$Li target there are the following 1,2,3 stable particles [t, t, p], [$^3$He, t, n] in two reactions γ+$^7$Li:

For excited $^7$Li target can be observed:

cluster+particle: ($^6$He+p), ($^6$Li+n)

two clusters ($^3$He+$^4$H), ($^4$He+t).

For excited isotopes products in the reaction γ+$^7$Li:

cluster+particle: $^4$H (t+n)

two clusters $^6$He (t+t)

The number of cases: "cluster + particle" is equal to three and for two clusters it is equal to three, the total sum - six.

Altogether (for four reactions) we have 16 cluster structures for nine isotopes:

$^3$H, $^4$H, $^3$He, $^4$He, $^5$He, $^6$He, $^5$Li, $^6$Li and $^7$Li: nine structures with one cluster + one particle and seven structures with two clusters .

Actually, the method based on the Dalitz diagram analysis is more informative than the presented one (see 3.3). For example, in the case of the excited state of the $^6$Li target, we have the cluster structure ($^3$He+t) where t(dn). Dalitz diagrams give additionally also one structure for ($^3$He+t) system: [(d+p)*+t]. In the case of $^7$Li target additionally for the structure ($^4$He+t) where $^4$He($^3$He,n) ,

we have structure [(t+p)*+t]. Thus, the Dalitz diagrams can show all versions, which can be useful in comparison with different theoretical models.

**4. Conditions necessary for the realization of the cluster structure study**

As it is shown, the above-mentioned method can be realized experimentally by means of the registration and measurement of full-parameters for two reaction products in three-body process. It is necessary to model all the experimental process for the choice of the type of detectors, the increase of solid angle of the registration apparatus, the provision of the accuracy of the reaction parameters, the determination of the expected number of events in the processes, etc.

The proposed program is presented in the following paragraphs:

Kinematics of the investigated processes:

investigation of the reaction kinematics including: determination of the particle and cluster parameters in the selected three-body photodisintegration reactions on $^6$Li, $^7$Li targets; estimation of contributions of the clusters and background reaction, calculation of the phase space and Dalitz diagrams, spectra of the effective masses for particule system.

Modeling of the nuclear photodisintegration process:

determination of the necessary parameter accuracy for the investigated processes and experimental conditions for the achievement of that precision; determination of the kinematics region of the investigated processes and expected yields (number of the events).

Choice of the parameters and the type of detectors:

providing the parameters of detectors and apparatus in accordance with the results of modeling the investigated reactions; providing the identification of the produced particle-clusters (p, d, t, $^3$He) in the investigated four reactions, the determination of kinetic energy and emitting angles with the necessary accuracy.

At first, it is supposed to carry out the experiment on the investigation of the reaction $\gamma+^7$Li→ t+t+p (applying the proposed method). The experiment should be realized on the bremsstrahlung beam of Yerevan Synchrotron at energy $E_\gamma$ =50-250 MeV (later also with polarized photons). It is supposed to investigate cluster structures of the excited states of the produced isotopes $^6$He, $^4$He and of $^7$Li target, in the reaction $\gamma+^7$Li→t+t+p in coincidence of the two tritons (see Fig.). It will also enable to investigate the two-body reactions:

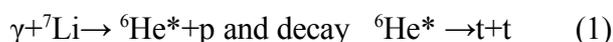
$\gamma+^7$Li→ $^6$He*+p and decay   $^6$He* →t+t     (1)

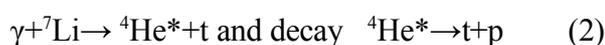
$\gamma+^7$Li→ $^4$He*+t and decay   $^4$He*→t+p     (2)

The reaction (2) will enable to compare the properties of the cluster structure of $^7$Li ($^4$He*+t) with the analogous structure, obtained in [10] in two-body reaction $\gamma + ^7$Li$\rightarrow ^4$He+t.

In the figure we present the Dalitz diagram of three-body photodisintegration reaction
$\gamma + ^7$Li $\rightarrow$ t +t +p at the incident energy $E_\gamma$ = 75 MeV.
The diagram corresponds to the kinetics energy $T_t$ in the CM system ($\gamma + ^7$Li), for two tritons (t) registered in coincidence. The lines (a) and (b) indicated in the diagram correspond to the excited state of the isotope $^6$He* decaying in two tritons in the reaction $\gamma + ^7$Li $\rightarrow ^6$He* +p ($^6$He* $\rightarrow$ t$_1$ +t$_2$). The line (a) is for the excited state $E_x$ = 20.9 ±0.3 MeV and the line(b) is for the excited state $E_x$ = 31.1 ± 1 MeV, as in the indicated experiment [7].

## Conclusion

We propose to investigate the cluster structure of light nuclei in three-body photonuclear spallation reactions on targets $^6$Li, $^7$Li (mainly two clusters and one nucleon) using bremsstrahlung and linear polarized photon beams. In these reactions the three parameters of each selected two particles from four charged particles (p, d, t, $^3$He) will be measured in the regime of the double coincidence (the third particle is nucleon). This enables to restore, for each reaction event, the energy of the incident photon beam. The analysis of the three-body processes is based on the application of the Dalitz diagram. In four photodisintegration reactions on the mentioned targets 16 cluster structures of 9 isotopes $^3$H, $^4$H, $^3$He, $^4$He, $^5$He, $^6$He, $^5$Li, $^6$Li and $^7$Li can be investigated.

- 9 structures with one cluster and one nucleon in isotopes: $^3$H (d+n); $^4$H(t+n); $^3$He (d+p);

    $^4$He [(t+p),($^3$He+n)]; $^6$Li [($^5$He+p), ($^5$Li+n)]; $^7$Li [($^6$He+p), ($^6$Li+n)].

- 7 structures with two clusters in isotopes: $^5$He (t+d) ; $^6$He (t+t); $^5$Li ($^3$He+d);

    $^6$Li [($^3$He+t), ( $^4$He +d)] ; $^7$Li [($^4$H+$^3$He), ($^4$He+t)]

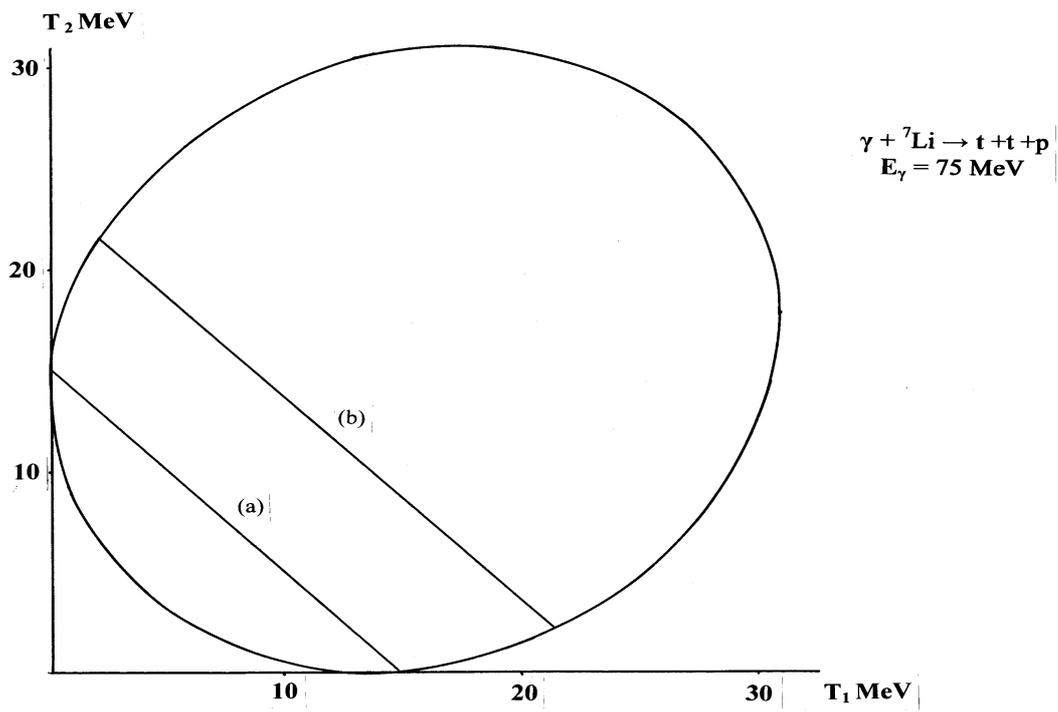

Fig.1. Dalitz diagram of three-body photodisintegration reaction $\gamma + {}^7\text{Li} \rightarrow t + t + p$ at the incident energy $E_\gamma = 75$ MeV (see text).